\documentclass[fleqn,10pt]{wlscirep}
\usepackage[utf8]{inputenc}
\usepackage[T1]{fontenc}
\usepackage{amsmath}
\usepackage{enumitem}
\usepackage{amssymb}
\usepackage{bbold}

\usepackage{color}
\usepackage{lipsum}

\newcommand{\sign}{\text{sign}}

\title{Probability density functions for photon propagation in a binary (isotropic-Poisson) statistical mixture
with unmatched positives/negatives refractive indexes}

\author[1,*]{Tiziano Binzoni}
\author[2]{Alain Mazzolo}

\affil[1]{Department of Radiology and Medical Informatics, University Hospital, Geneva, 1211, Switzerland}
\affil[2]{Universit\'e Paris-Saclay, CEA, Service d'\'Etudes des R\'eacteurs et de Math\'ematiques Appliqu\'ees, 91191, Gif-sur-Yvette, France}

\affil[*]{tiziano.binzoni@unige.ch}

\begin{abstract}
The exact homogenized probability density function, for a photon making a step of length $s$ has been analytically derived for  
a binary (isotropic-Poisson) statistical mixture with unmatched refractive indexes.
The companions, exact, homogenized probability density function for a photon to change direction (``scatter'') with an angle $\vartheta$,
and the homogenized albedo, have also been obtained analytically.
These functions also hold even in the case of negative  refractive indexes  and 
allow one to reduce hundreds of MC simulations of photon 
propagation in complex binary (isotropic-Poisson) statistical mixtures, 
to only one MC simulation, for an equivalent homogeneous medium.
Note, that this is not an approximate approach, but a mathematically equivalent and exact result.
Additionally, some tutorial examples of homogenized MC simulations are also given.
\end{abstract}

\begin{document}

\flushbottom
\maketitle
\thispagestyle{empty}

\section*{Introduction}
\label{}

Monte Carlo (MC) simulations of heterogeneous and disordered media 
represent an important subject of study, because the underlying theory 
may be at the basis of many physical and natural systems
\cite{ref:Barthelemy2008,ref:Svensson2013,ref:Svensson2014,ref:DavisMarshak2004,ref:kostinskishaw2001,
ref:Malvagi1993,ref:LarsenVasques2011,ref:Zuchuat1994,ref:Haran2000,Ref:Mercadier2009}. 
Due to their complex geometries and statistical properties, these media are very difficult to tackle, in particular, with analytical techniques. 
For this reason, numerical MC simulations represent a ``gold standard'' allowing to solve many problems in this domain.

Among these media, binary (isotropic-Poisson) statistical mixtures are of particular interest,
because they allow us to develop algorithms permitting to treat them as they would be homogeneous media. 
This equivalence greatly simplifies MC simulations, since it permits to  reduce hundreds of independent MC simulations (the classical approach), 
necessary to compute the means of the wanted parameters,
to just one MC simulation (the present approach)\cite{ref:Pomraning1991}.

It is in this framework that, in a previous work, we were able to calculate the homogenized probability density function (pdf) for a photon 
to make a single step of length $s$, in a binary (isotropic-Poisson) statistical mixture\cite{ref:Binzoni2023}.
This allows us to treat the medium as homogeneous and the pdf plays the role of the celebrated Beer-Lambert-Bouguer law in classical MC simulations.
Note that this is not an approximate approach to the MC simulations, applied to the binary (isotropic-Poisson) statistical mixtures,
but rather a mathematically {\it exact equivalence}. 

One limitation of the results presented in Ref.~\cite{ref:Binzoni2023} is that the binary medium was treated as
if the different bunch of materials composing it had the same refractive index, which is often not the case in reality.
Moreover, the homogenized phase function, describing the changes in photons direction after a scattering event,
was not assessed at that time. 
This is a non-trivial points to be considered, if moreover  we want  a medium composed with materials of different refractive indexes. In fact, in this case the photons can also ``scatter'' on the surfaces separating the different materials, rendering the exact determination of the homogenized phase function more complex. 

This is why, the above-mentioned reasons prompted us to treat in the present contribution the following points:
1) Consider a binary (isotropic-Poisson) statistical mixtures
with unmatched (different) refractive indexes; 
2) Derive the homogenized pdf for a photon to make a step of length $s$ (SSF: single step function);
3) Derive the homogenized pdf for a photon to change direction (``scatter'') with an angle $\vartheta$ (PF: phase function); 
4) For the sake of completeness and, to the best of our knowledge, the lack of literature on the topic, 
the possibility in the previous 3 points to have negative refractive indexes,
is also considered.

To facilitate the reading,  we will carry on use the word ``homogenized'' to clarify that, e.g., a function or a parameter
are the result of the homogenization procedure, and ``classical'' when they must be interpreted in the usual sense.
If not explicitly stated, the context should be enough to clarify this point.

\section*{Theory}
\label{Theory}

\subsection*{The medium}

The characteristics of the medium considered in the present contribution --- a binary (isotropic-Poisson) statistical mixture ---  
have already been presented in other works \cite{ref:Larmier2016,ref:Larmier2018a,ref:Larmier2018b,ref:Binzoni2023}.
For this reason, we will only provide a brief summary of the main points necessary to understand the theory presented in the following sections.

Intuitively, a realization of the random 3D medium, i.e., one among many possibilities,  is obtained by:
\begin{enumerate}
\item
Generating a set of planes with a uniformly distributed random orientation.
This is done is such a way that an arbitrary  line intersecting the random planes is divided in $l_i$ random segments (see Fig. \ref{fig_tessels-crop};
dashed arrow A) 
\begin{figure}[htbp]
\includegraphics[width=9cm]{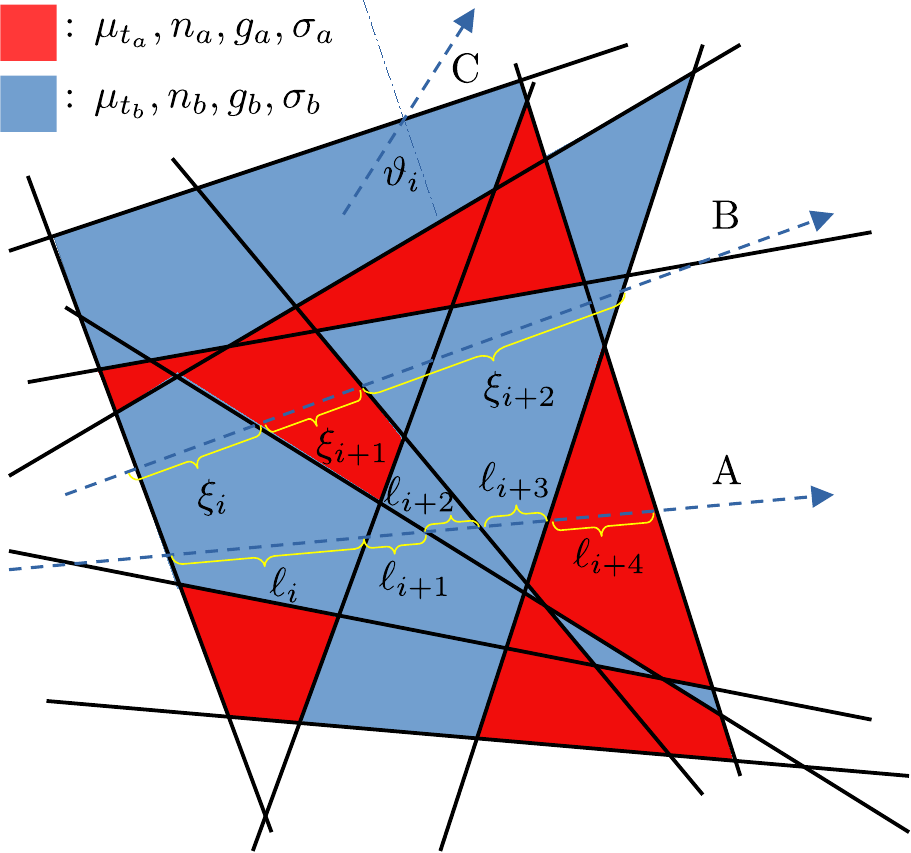}
\centering
\caption{Simplified 2D intuitive representation of a realization of a binary (isotropic-Poisson) statistical mixture.
The figure  extends to the infinite plane; for obvious reasons, here only a small region is shown.
Continuous black lines represent the planes. 
Colored regions represent the tessels of type a and b.
The dash-dotted line, near label C, represents the normal to the plane.
For the dashed arrows A, B and C see text.
}\label{fig_tessels-crop}
\end{figure}
of lengths described by the probability density function (pdf).
\begin{align}
p_L(l)=\frac{1}{L}e^{-\frac{l}{L}}.
\label{eq:pL}
\end{align}
Thus, when 
an arbitrary
 line intersects the planes, the generated incident angles, $\vartheta_i$,  (e.g., Fig. \ref{fig_tessels-crop};
dashed arrow C) 
are uniformly distributed over $[0,\frac{\pi}{2}]$, i.e., with pdf
\begin{align}
p_{\vartheta_i}(\vartheta_i)=\frac{2}{\pi}\mathbb{1}_{\left[0,\frac{\pi}{2}\right]}(\vartheta_i),
\label{eq:pthetai0pid2}
\end{align}
where $\mathbb{1}_{A}(x)$ denotes the indicator function
\begin{eqnarray}
  \mathbb{1}_{A}(x) 
 = 
 \left\lbrace
  \begin{array}{lll}
    1  
    &~\mathrm{if~~} x \in A
    \\
    0
    &~\mathrm{if~~} x \notin A  \, .
  \end{array}
\right.
  \label{driftstarsurvivalinftyfinitedelta0}
\end{eqnarray}
\item
To obtain a realization of the  binary statistical mixture, the different 3D regions, delimited by the planes, are randomly ``colored'' with two colors,
i.e.;  with  probability $1-P$ to be ``red'' and a probability P to be ``sky blue'' (Fig. \ref{fig_tessels-crop}).
This will generate ``red'' and ``sky blue'' volumes called, in the present context,  ``tessels''.
As a consequence,
an arbitrary  line intersecting the tessels (see Fig. \ref{fig_tessels-crop};
dashed arrow B) will be divided into $\xi_i$ random segments with pdf
\begin{equation}
p_{Tes}(\xi;\sigma)=\sigma e^{-\sigma \xi}; \qquad  \sigma \in \{\sigma_a,\sigma_b\},
\label{eq:pdfTesselLength}
\end{equation}
where 
\begin{align}
\sigma_a=\frac{1}{L}P; \qquad \sigma_b=\frac{1}{L}(1-P).
\label{eq:sigmasigma4}
\end{align}
From now on, parameters referring to ``red'' and ``sky blue'' tessels will be labeled with ``a'' and  ``b'', respectively.
\item
The optical parameters will then be linked to the tessels of type a and b, i.e.:
$\mu_{a_i}$, absorption coefficient; $\mu_{s_i}$,  scattering coefficient; $\mu_{t_i}=\mu_{a_i}+\mu_{s_i}$ extinction coefficient; 
$n_i$, refractive index; $g_i$, mean cosine (see below);  with $i \in \{a,b\}$.
\end{enumerate}

Note that in the present contribution {\it we will always consider $|n_a|>|n_b|$}. 
This does not affect the generality of the results. 

\subsection*{Homogenized Single step function $p_{ s_{\rm mix}}(.)$}

We  will now derive  the homogenized  pdf $p_{ s_{\rm mix}}(.)$ for a photon to make a single step of length $s$, i.e.; the homogenized SSF.
In this case, we have three possibilities for the photon: 
1) it reaches a distance $s$ inside the tessel at the point where it is absorbed; 
2) it reaches a distance $s$ inside the tessel at the point where it is scattered;
3) it reaches the tessel boundary where it is reflected/refracted, because $n_a \neq n_b$
(note that this can also be seen as a kind of ``scattering'' since the photon changes direction).

In a previous contribution \cite{ref:Binzoni2023} we have demonstrated that:
\begin{itemize}
\item
The homogenized pdf for a photon to make a step of length $s$,
remaining inside a tessel of type $i\in \{a,b\}$ (absorbed or scattered), is $(\mu_{t_i}+\sigma_i)e^{-(\mu_{t_i}+\sigma_i)s}$.
\item
The homogenized pdf to make a step that falls on the tessel boundary (reflected/refracted), of type $i\in \{a,b\}$, at distance $s$, is also $(\mu_{t_i}+\sigma_i)e^{-(\mu_{t_i}+\sigma_i)s}$.
\item
The probability for a photon to remain inside the tessel $i$, after a single step, is
\begin{align}
P_0(\mu_{t_i},\sigma_i)=\frac{\mu_{t_i}}{\mu_{t_i}+\sigma_i}; \qquad i\in \{a,b\}.
\label{eq:P0}
\end{align}
\end{itemize}
These results allow us to express in general (i.e.; 
simultaneously taking
into account  absorption, scattering and reflection/refraction)
the pdf for a photon to make a step of length $s$ in a tessel of type $i$ as
\begin{align}
p_{ s}(s;\mu_{t_i} ,\sigma_i )=
P_0(\mu_{t_i},\sigma_i)(\mu_{t_i}+\sigma_i)e^{-(\mu_{t_i}+\sigma_i)s}+
[1-P_0(\mu_{t_i},\sigma_i)](\mu_{t_i}+\sigma_i)e^{-(\mu_{t_i}+\sigma_i)s}
=
 (\mu_{t_i}+\sigma_i)e^{-(\mu_{t_i}+\sigma_i)s}.
\label{eq:psFINAL}
\end{align}
Thus, the pdf $p_{ s_{\rm mix}}(.)$ can be written as
\begin{align}
p_{s_{\rm mix}}(s;\mu_{t_a},\mu_{t_b},\sigma_a,\sigma_b,P)
=(1-P) p_{ s}(s;\mu_{t_a},\sigma_a)+P p_{ s}(s;\mu_{t_b},\sigma_b),
\label{eq:psFINALmix}
\end{align}
where $1-P$ and $P$ are the probabilities to have a tessel of type a or b, respectively.
Finally, by substituting Eqs. (\ref{eq:P0}) and (\ref{eq:psFINAL}) into Eq.~(\ref{eq:psFINALmix}), we obtain
the desired homogenized SSF
\begin{align}
p_{s_{\rm mix}}(s;\mu_{t_a},\mu_{t_b},L,P)
=
(1-P)\left(\mu_{t_a}+\frac{P}{L}\right)e^{-\left(\mu_{t_a}+\frac{P}{L}\right)s}
+
P\left(\mu_{t_b}+\frac{1-P}{L}\right)e^{-\left(\mu_{t_b}+\frac{1-P}{L}\right)s}.
\label{eq:psFINALmixLP}
\end{align}
Given that  $p_{s_{\rm mix}}(s;.)$ is a pdf, we have $\int_0^{+\infty}p_{s_{\rm mix}}(s;.)ds=1$. 

As already explained in references \cite{ref:dEon2018,ref:BinzoniMartelli2022}, if the photon starts from a ``fixed'' boundary, e.g., from a fixed light source or from the 
surface that delimits the medium, the homogenized pdf to make a step of length $s$ can be obtained as
\begin{align}
p_{s_{\rm fix}}(s;\mu_{t_a},\mu_{t_b}L,P)&=
\frac{1-\int_0^{s} p_{s_{\rm mix}}(s';\mu_{t_a},\mu_{t_b},L,P)ds'}
{\int_0^{+\infty} s' p_{s_{\rm mix}}(s';\mu_{t_a},\mu_{t_b},L,P)ds'}
\nonumber \\
&=\frac{\left(\mu_{t_a}+\frac{P}{L}\right)\left(\mu_{t_b}+\frac{1-P}{L}\right)\left[(1-P)e^{-\left(\mu_{t_a}+\frac{P}{L}\right)s}+Pe^{-\left(\mu_{t_b}+\frac{1-P}{L}\right)s}\right]
 L}
{ L\left[P  \mu_{t_a}+(1-P)  \mu_{t_b}\right] + 2 P^2 - 2 P + 1}.
\label{eq:uncorr}
\end{align}
Even in this case, we obviously get $\int_0^{+\infty}p_{s_{\rm fix}}(s;.)ds=1$.
The validity of  $p_{s_{\rm mix}}(.)$ [Eq.~(\ref{eq:psFINALmixLP})] and, consequently, $p_{s_{\rm fix}}(.)$ [Eq.~(\ref{eq:uncorr})] has been tested with  {\it ad hoc}
``gold standard'' classical MC simulations (see some examples below).

\subsection*{Homogenized phase function $p_{\Theta}(\vartheta)$}

\begin{figure*}[htbp]
\includegraphics[width=15cm]{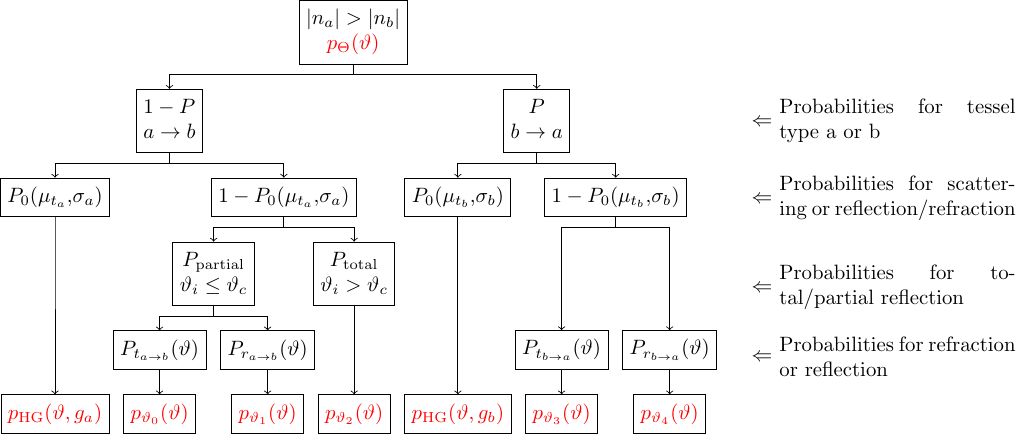}
\centering
\caption{Schematic summarizing the main steps allowing to derive the PF $p_{\Theta}(\vartheta)$.}
\label{treeSchem2}
\end{figure*}

The derivation of the homogenized PF $p_{\Theta}(\vartheta)$, {\it for positive and negative refractive indexes}, is more complex than that of the SSF. 
For this reason, we will first summarize the whole procedure with the help of a schematic.
At the top of Fig. \ref{treeSchem2}, 
$p_{\Theta}(\vartheta)$ represents the wanted PF, where 
$\vartheta$ (scattering angle)  is the change in the photon direction after a ''scattering'' event.
As mentioned in the previous section, we will also consider  reflection/refraction as ''scattering'' events,
due to the observed induced  changes in the photon's direction.

Thus, to derive  $p_{\Theta}(\vartheta)$, we must 
answer the following questions:
\begin{enumerate}
\item
Is the scattering event  due 
to a photon that comes from a tessel a or from b?
This choice is made by exploiting the  fact that the photon has probability $1-P$ to be in a, and $P$ to be in b
(Fig. \ref{treeSchem2}; 2$^{\rm nd}$ line). 
\item
Is the photon scattered inside 
the tessel or  is it 
reflected/refracted on the tessel boundary
(Fig. \ref{treeSchem2}; 3$^{\rm rd}$ line)?
This choice is made by knowing the probability $P_0(\mu_{t_i},\sigma_i)$ [Eq.~(\ref{eq:P0})] 
for a photon to remain inside the tessel  $i\in \{a,b\}$  after a single step.
\item
If there is a reflection, 
is it total or a partial
(Fig. \ref{treeSchem2}; 4$^{\rm th}$ line)?
This choice is made by knowing the probability to have a partial reflection $P_{\rm partial}$; derived thanks to the critical angle $\vartheta_c$.
Obviously, if the photon goes from a tessel of type b to a, this case does not exist 
(because when going from a  low to a higher refractive index there is no critical angle).
\item
Does the photon actually
 undergo
a reflection or a refraction (Fig. \ref{treeSchem2}; 5$^{\rm th}$ line)?
This choice is made by using the probabilities  $P_{t_{a \rightarrow b}}(\vartheta)$, $P_{r_{a \rightarrow b}}(\vartheta)$,
$P_{t_{b \rightarrow a}}(\vartheta)$ and $P_{r_{b \rightarrow a}}(\vartheta)$ that will be derived in the following sections.
Note the dependence of the scattering angle.
\end{enumerate}
Finally, at the end of each tree branch of Fig. \ref{treeSchem2} (6$^{\rm th}$ line), we can derive the relatives pdfs,
taking into account  for the different physical conditions described in points 1--4, and
allowing to build  the homogenized $p_{\Theta}(\vartheta)$, i.e.;

\begin{align}
p_{\Theta}(\vartheta)&=\left(1-P \right) \times
\nonumber \\
&\Bigl\{P_0(\mu_{t_a},\sigma_a)p_{\rm HG}(\vartheta,g_a)
 +
\left(1-P_0(\mu_{t_a},\sigma_a)\right)
\left[P_{\rm partial}(P_{t_{a\rightarrow b}}(\vartheta)p_{\vartheta_0}(\vartheta)+ P_{r_{a\rightarrow b}}(\vartheta) p_{\vartheta_1}(\vartheta))
+ P_{\rm total} p_{\vartheta_2}(\vartheta)\right]\Bigr\}
\nonumber \\
&+
 \phantom{aaa} P  \phantom{aa}\times
\nonumber \\
&\Bigl\{P_0(\mu_{t_b},\sigma_b)p_{\rm HG}(\vartheta,g_b)\Bigr.
+
\left(1-P_0(\mu_{t_b},\sigma_b)\right)
\left[P_{t_{b\rightarrow a}}(\vartheta)p_{\vartheta_3}(\vartheta)+P_{r_{b\rightarrow a}}(\vartheta)p_{\vartheta_4}(\vartheta)\right] \Bigr\}.
\label{eq:pthetavattheta}
\end{align}
The validity of  $p_{\Theta}(\vartheta)$ [Eq.~(\ref{eq:pthetavattheta})]   has been tested with  {\it ad hoc}
``gold standard'' classical MC simulations (see some examples below).
To avoid a too heavy formalism, the refractive indexes will never appear as variables in the functions 
composing  Eq. (\ref{eq:pthetavattheta}). 
The context will be sufficient to clarify this point.
The parameters $\mu_{t_i}$ and $\sigma_i$, $i \in \{a,b\}$, will appear  only when necessary.

\subsubsection*{Probability density function $p_{\rm HG}(\vartheta,g_i)$}

\begin{figure}[htbp]
\includegraphics[width=3.5cm]{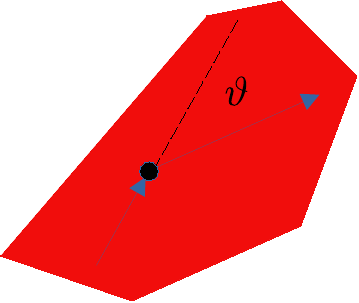}
\centering
\caption{Example of the change in direction $\vartheta$ of a photon scattering on a scatterer located inside a  tessel.}\label{HG-crop}
\end{figure}

For the pdf of a photon to have a scattering angle $\vartheta$, 
due to
a scattering event inside a tessel $i$, we
will use the celebrated Henyey-Greenstein function\cite{ref:HenyeyGreenstein1941,ref:Binzoni2006}, applicable in a large number of situations.
\begin{align}
p_{\rm HG}(\vartheta,g_i)=\frac{1}{2}\frac{(1-g_i^2)\sin(\vartheta)}{\left[1+g_i^2-2 g_i \cos(\vartheta)\right]^{\frac{3}{2}}};
\qquad i\in \{a,b\}
\label{eq:HenyGree}
\end{align}
where $\vartheta \in [0,\pi]$ (Fig. \ref{HG-crop}), 
and $g_i$ is the mean cosine.
However, other  functions can  obviously be used, depending on the particular needs of the reader. 

We have seen above  that the probability for a photon to occur in this type of scattering event, described by  $p_{ HG}(\vartheta,g_i)$, is
$P_0(\mu_{t_i},\sigma_i)$ [Eq.~(\ref{eq:P0})].

\subsubsection*{Probability density function  $p_{\vartheta_0}(\vartheta)$ (partial refraction: $a \rightarrow b$, $\vartheta_i\le\vartheta_c$)}

In the present and in the  following sections, we will use the fact that\cite{ref:Veselago1968}:
\begin{itemize}
\item
The well-known Snell's equation 
\begin{align}
n_a \sin( \vartheta_i) = n_b \sin(\vartheta_t),
\label{snell12}
\end{align}
holds, in general, for positive or negative refractive indexes.
The practical consequence is that if {\it one} of the two refractive indexes is negative, then the angle of refraction $\vartheta_t$ simply becomes
$-\vartheta_t$.
\item
In the case of  partial refraction, the critical angle is expressed as
\begin{align}
\vartheta_c=\arcsin\left(\left|\frac{n_b}{n_a}\right|\right).
\end{align}
Note that in the present contribution we can have partial refraction only when we go from a$\rightarrow$b, because  
we  have imposed $|n_a|>|n_b|$.
\item
If the photon goes from b$\rightarrow$a, Snell's equation must be written as
\begin{align}
n_b \sin( \vartheta_i) = n_a \sin(\vartheta_t)
\label{snell12ba}
\end{align}
\end{itemize}
(no critical angle in this case).

\begin{figure}[htbp]
\includegraphics[width=12cm]{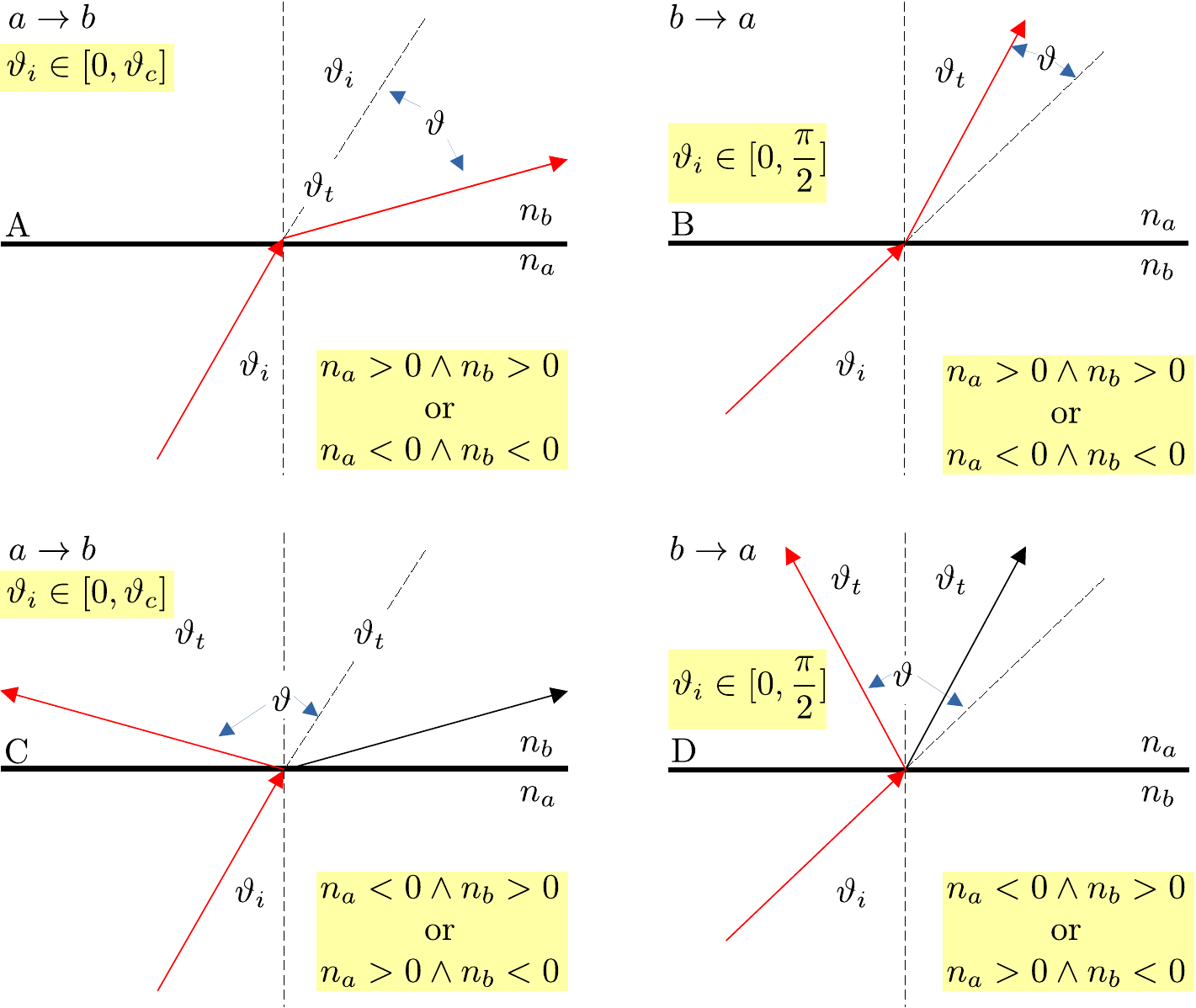}
\centering
\caption{Four different cases of refraction. A and C are related to the pdf $p_{\vartheta_0}(\vartheta)$. 
B and D are related to the pdf $p_{\vartheta_3}(\vartheta)$}.
\label{SnellTheta-crop}
\end{figure}

Thus, in this section we will derive the pdf $p_{\vartheta_0}(\vartheta)$ in the case of partial refraction, i.e.; when
the random angle $\vartheta_i\le\vartheta_c$ and the photon goes from a$\rightarrow$b.
For explanatory reasons, we will treat before the case where the refractive indexes are simultaneously positive or negative. 
This case is schematically represented in Fig.~\ref{SnellTheta-crop}A.

Considered that $\vartheta_i$ is uniformly distributed over $[0,\frac{\pi}{2}]$ [Eq. (\ref{eq:pthetai0pid2})],
the probability to have a refraction/reflection  is
\begin{align}
P_{\rm partial}=\frac{\vartheta_c}{\pi/2}.
\label{eq:Ppartial15}
\end{align}

Now, in the case of refraction,  the relation that links  $\vartheta_t$ and $\vartheta$ is
\begin{align}
\vartheta_t=\vartheta_i+\vartheta.
\label{eq:thetap0}
\end{align}
By substituting Eq. (\ref{eq:thetap0}) into Eq. (\ref{snell12}), and by solving for $\vartheta$, we can express
$\vartheta$ as a function of $\vartheta_i$.
\begin{align}
\vartheta   = \arcsin\left(\frac{n_a}{n_b}\sin(\vartheta_i)\right)-\vartheta_i \equiv g(\vartheta_i),
\label{eq:gvartheta17}
\end{align}
and
\begin{align}
   \frac{d g(\vartheta)}{d\vartheta} = \frac{n_a}{n_b} \frac{\cos(\vartheta)}{\sqrt{1-\frac{n_a^2}{n_b^2} \sin ^2(\vartheta )}} - 1.
\label{eq:Dgvartheta17}
\end{align}
By considering  two cases, for 
$\vartheta_i=0$ and $\vartheta_i=\vartheta_c$, Eq. (\ref{eq:gvartheta17}) tells us that
$\vartheta$ 
\begin{align}
\vartheta   \in [0,\frac{\pi}{2} -\vartheta_c].
\end{align}
From Eq.~(\ref{eq:Dgvartheta17})
we can see that the equation $dg(\vartheta)/d\vartheta = 0$ has no real solutions, therefore, $dg(\vartheta)/d\vartheta$ is either positive or negative and $g(\vartheta)$ is an injective function over the interval $[0,\frac{\pi}{2} -\vartheta_c]$. 
As a result, $p_{\vartheta_0}(\vartheta)$ can be derived over this interval using the classical method\cite{ref:Hsu2020}
\begin{align}
p_{\vartheta_0}(\vartheta)=p_{\vartheta_c}\left(g^{-1}(\vartheta)\right)\left|\frac{d g^{-1}(\vartheta)}{d \vartheta}\right|,
\label{eq:pthetatheo}
\end{align}
where
\begin{align}
p_{\vartheta_c}(\vartheta_i)=\frac{1}{\vartheta_c} \mathbb{1}_{[0,\vartheta_c]}(\vartheta_i)
\label{eq:pthetaccc}
\end{align}
is the canonical pdf of $\vartheta_i$ over $[0,\vartheta_c]$,
\begin{align}
g^{-1}(\vartheta )=\vartheta_i=\arccos\left(  \frac{n_a-  n_b \cos(\vartheta)}{\left(n_a^2 +n_b^2 -2 n_a n_b \cos(\vartheta)\right)^{\frac{1}{2}}}\right)
\label{eq:gmenouno21}
\end{align}
(from Eq. (\ref{eq:gvartheta17})) and
\begin{align}
\frac{d g^{-1}(\vartheta)}{d \vartheta}=  \frac{n_a n_b \cos(\vartheta)-n_b^2}{n_a^2+n_b^2-2 n_a n_b \cos(\vartheta)}.
\label{eq:dergmenouno21}
\end{align}
Finally, by substituting Eqs. (\ref{eq:pthetaccc}), (\ref{eq:gmenouno21}) and (\ref{eq:dergmenouno21}) into Eq. (\ref{eq:pthetatheo})
we get
\begin{align}
p_{\vartheta_0}(\vartheta)=
\frac{1}{\vartheta_c} \mathbb{1}_{[0,\frac{\pi}{2}-\vartheta_c]}(\vartheta)
\left|  \frac{n_a n_b \cos(\vartheta)-n_b^2}{n_a^2+n_b^2-2 n_a n_b \cos(\vartheta)} \right| .
\label{eq:ptheta}
\end{align}
It is maybe useful to highlight again the fact that the case where both $n_a$ and $n_b$ are negative is equivalent to the case where both $n_a$ and $n_b$ are positive.

The last remaining step is to express $p_{\vartheta_0}(\vartheta)$ in the case where one of the refractive indexes is negative.
This condition implies that the relationship existing between the incident and the refracted angle is (Fig.~\ref{SnellTheta-crop}C)
\begin{align}
-\theta_t=\theta_i-\vartheta.
\label{eq:menothetat24}
\end{align}
By using Eq. (\ref{eq:menothetat24}) in place of Eq. (\ref{eq:thetap0}), and by repeating the above procedure, we obtain
\begin{align}
p_{\vartheta_0}(\vartheta)=
\frac{1}{\vartheta_c} \mathbb{1}_{[0,\frac{\pi}{2}-\sign(n_a n_b) \vartheta_c]}(\vartheta)
\left|  \frac{n_a n_b \cos(\vartheta)-n_b^2}{n_a^2+n_b^2-2 n_a n_b \cos(\vartheta)} \right| .
\label{eq:pthetafinal}
\end{align}
It is interesting to note that Eq. (\ref{eq:pthetafinal}) also includes Eq.~(\ref{eq:ptheta}).
For this reason,  Eq.~(\ref{eq:pthetafinal}) will be adopted as a final   model for $p_{\vartheta_0}(\vartheta)$.

We have seen that the probability to have a reflection or a refraction is $P_{\rm partial}$ [Eq. (\ref{eq:Ppartial15})],
but Eq.~(\ref{eq:pthetavattheta}) tells us that we also need the probability to have only a refraction $P_{t_{a \rightarrow b}}(\vartheta)$, 
i.e.; the probability to have $p_{\vartheta_0}(\vartheta)$.  
$P_{t_{a \rightarrow b}}(\vartheta)$ is classically given by the Fresnel's formula for non-polarized light\cite{ref:BornWolf1986,ref:Hecht2012},
but where in the present case the angles of incidence and refraction must be expressed in terms of $\vartheta$.
So, starting from the Fresnel's formula and substituting Eqs. (\ref{eq:thetap0}) and (\ref{eq:gmenouno21}), we can write
\begin{align}
P_{t_{a \rightarrow b}}(\vartheta) &= 1-\frac{1}{2}\left[ \frac{\sin^2(\vartheta_i-\vartheta_t)}{\sin^2(\vartheta_i+\vartheta_t)} +  
\frac{\tan^2(\vartheta_i-\vartheta_t)}{\tan^2(\vartheta_i+\vartheta_t)} \right]
\nonumber \\
&=
1- \frac{1}{2}\left[ \frac{\sin^2(-\vartheta)}{\sin^2(2 \vartheta_i+\vartheta )} +  
\frac{\tan^2(-\vartheta)}{\tan^2(2 \vartheta_i+\vartheta )} \right]
\nonumber \\
&=
1-\frac{1}{2}\left[
\csc\left(\vartheta + 2 \arccos\left(\frac{n_a-  n_b \cos(\vartheta)}{\left(n_a^2 +n_b^2 -2 n_a n_b \cos(\vartheta)\right)^{\frac{1}{2}}}\right)\right)^2\sin(\vartheta)^2
\right.
\nonumber \\
&+
\left. 
\cot\left(\vartheta + 2 \arccos\left(\frac{n_a-  n_b \cos(\vartheta)}{\left(n_a^2 +n_b^2 -2 n_a n_b \cos(\vartheta)\right)^{\frac{1}{2}}}\right)\right)^2\tan(\vartheta)^2
\right].
\label{eq:Pt}
\end{align}
Finally, the case for negative refractive indexes ($n_a$ or $n_b$) can be obtained by following the same approach, but by using Eq. (\ref{eq:menothetat24}) instead of  Eq. (\ref{eq:thetap0}), i.e.;
\begin{align}
P_{t_{a \rightarrow b}}(\vartheta) = 
1-\frac{1}{2}&\left\{
\left[\csc\left(\vartheta + 2 \arccos\left(\frac{n_a-  n_b \cos(\vartheta)}{\left(n_a^2 +n_b^2 -2 n_a n_b \cos(\vartheta)\right)^{\frac{1}{2}}}\right)\right) \sin(\vartheta)\right]^{2 \, \sign(n_a n_b)}
\right.
\nonumber \\
&+
\left. 
\left[\cot\left(\vartheta + 2 \arccos\left(\frac{n_a-  n_b \cos(\vartheta)}{\left(n_a^2 +n_b^2 -2 n_a n_b \cos(\vartheta)\right)^{\frac{1}{2}}}\right)\right) \tan(\vartheta)\right]^{2 \, \sign(n_a n_b)}
\right\}.
\label{eq:Ptfinal}
\end{align}
We can see that formally Eq. (\ref{eq:Ptfinal}) also includes Eq. (\ref{eq:Pt}), and this is why, from now on, we will only use Eq. (\ref{eq:Ptfinal}).

\subsubsection*{Probability density function  $p_{\vartheta_1}(\vartheta_1)$ (partial reflection: $a \rightarrow b$, $\vartheta_i\le\vartheta_c$)}

As in the previous section, the probability to have $\theta_i  \in [0,\vartheta_c]$, i.e., a reflection/refraction, obviously remains $P_{\rm partial}$ 
[Eq. (\ref{eq:Ppartial15})] .

\begin{figure}[htbp]
\includegraphics[width=6cm]{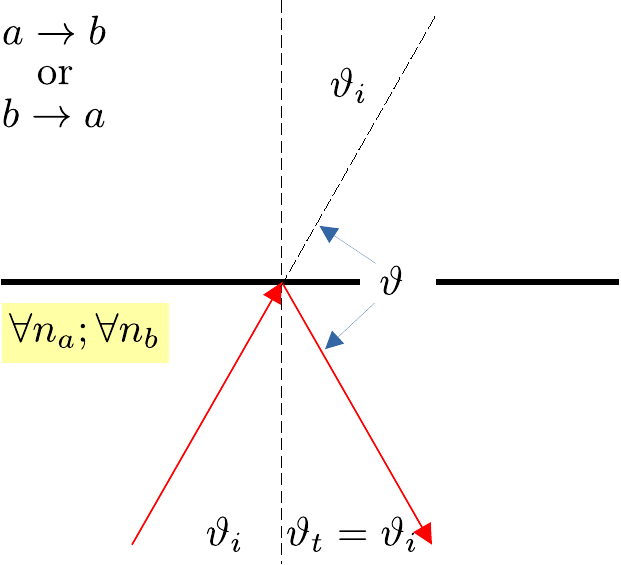}
\centering
\caption{This figure represents the reflection of a photon on a tessel boundary.
This holds in general for positive or negative refractive indexes.
However, if the photon goes from $a \rightarrow b$, then there is a critical angle $\vartheta_c$
for $\theta_i$ (not represented), while, 
if the photon goes from $b \rightarrow a$, there is no critical angle.
}\label{SnellTheta1-crop}
\end{figure}

So, in the case of reflection, Fig. \ref{SnellTheta1-crop} 
shows that 
\begin{align}
\vartheta  =\pi-2 \vartheta_i 
\label{eq:vartheta2828}
\end{align}
and, as a consequence,  $\vartheta$ is uniformly distributed over the interval
\begin{align}
\vartheta   \in [\pi-2 \vartheta_c,\pi].
\end{align}
This allows us to immediately express $p_{\vartheta_1}(\vartheta)$ as
\begin{align}
p_{\vartheta_1}(\vartheta) = \frac{1}{2\vartheta_c} \mathbb{1}_{[\pi-2 \vartheta_c,\pi]}(\vartheta) .
\label{eq:ptheta1}
\end{align}

Similarly to the previous section, we will exploit again Fresnel's formula to derive the probability $P_{r}(\vartheta)$ to have Eq.~(\ref{eq:ptheta1}), 
i.e.; the probability that the photon is reflected in the case $\vartheta_i\le\vartheta_c$.
By inserting Eq. (\ref{eq:vartheta2828}) into Snell's law
\begin{align}
\vartheta_t=\arcsin\left(\frac{n_a}{n_b}\sin(\frac{\pi-\vartheta}{2})\right),
\label{eq:varthetat3131}
\end{align}
and by inserting Eqs. (\ref{eq:vartheta2828})  and (\ref{eq:varthetat3131}) into Fresnel's formula for reflection, we obtain  
\begin{align}
P_{r_{a \rightarrow b}}(\vartheta)&=  \frac{1}{2}\left[ \frac{\sin^2(\vartheta_i-\vartheta_t)}{\sin^2(\vartheta_i+\vartheta_t)} +  
\frac{\tan^2(\vartheta_i-\vartheta_t)}{\tan^2(\vartheta_i+\vartheta_t)} \right]
\nonumber \\
&=
\frac{1}{2 n_b^2}
\left[n_b^2 + \cos(\vartheta) \left(  n_a^2-n_b^2 + n_a^2 \cos(\vartheta) \right) \right]
\nonumber \\
&\times
\cot\left( \frac{\vartheta}{2} +\arcsin\left( \left|\frac{n_a}{n_b}\right|\cos\left(\frac{\vartheta}{2}\right)  \right)\right)^2
\csc\left( \frac{\vartheta}{2} +\arccos\left( \left|\frac{n_a}{n_b}\right|\cos\left(\frac{\vartheta}{2}\right)  \right)\right)^2
\label{Eq:Prtheta1}
\end{align}
According to Fig.  \ref{SnellTheta1-crop} and e.g. Ref.\cite{ref:Veselago1968} we know that Eq. (\ref{Eq:Prtheta1}) must hold for positive or negative refractive indexes.
It is sufficient to  consider only their absolute values. This is why $|.|$ appears in the equation.

\subsubsection*{Probability density function  $p_{\theta_2}(\theta )$ (total reflection: $a \rightarrow b$, $\vartheta_i>\vartheta_c$)}

In the case where the photon goes from $a \rightarrow b$ and $\vartheta_i>\vartheta_c$, we always have total reflection.
Thus, using Eq. (\ref{eq:Ppartial15}), the probability to have this condition is
\begin{align}
P_{\rm total}=1-P_{\rm partial}.
\end{align}
In this case, the relationship between $\vartheta$ and $\vartheta_i$ (Fig. \ref{SnellTheta1-crop}) is still described by Eq. (\ref{eq:vartheta2828}),
however now $\vartheta_i$ is uniformly distributed over the interval
\begin{align}
\theta_i \in [\vartheta_c,\pi/2].
\end{align}
As a consequence, $\vartheta$ is uniformly distributed over
\begin{align}
\vartheta\in [0,\pi -2\vartheta_c].
\end{align}
This allows us to directly express the pdf $p_{\vartheta_2}(\vartheta)$ as
\begin{align}
p_{\vartheta_2}(\vartheta) = \frac{1}{\pi-2\vartheta_c} \mathbb{1}_{[0,\pi -2\vartheta_c]} (\vartheta) .
\label{eq:ptheta1a}
\end{align}

\subsubsection*{Probability density function  $p_{\vartheta_3}(\vartheta)$ (refraction: $b \rightarrow a$; no critical angle)}

Now, let us treat the case where the photon travels from $b \rightarrow a$, and where the refractive indexes are positive or negative (\ref{SnellTheta-crop}B). 
In this case, there is no critical angle,
and the pdf $p_{\vartheta_i}(\vartheta_i)$ is  expressed as in Eq. (\ref{eq:pthetai0pid2}).
Snell's equation is written as  reminded in Eq. (\ref{snell12ba}).
The relationship between the different angles is (Fig. \ref{SnellTheta-crop}B)
\begin{align}
\theta_t=\theta_i-\theta
\end{align}
By applying the same cumbersome procedure reported in Eqs. (\ref{eq:gvartheta17})--(\ref{eq:ptheta}),
we obtain
\begin{align}
p_{\vartheta_3}(\vartheta) = \frac{1}{\pi} \mathbb{1}_{[0,\frac{\pi}{2} - \vartheta_c]}(\vartheta)
\left|  1-\frac{n_b^2-n_a^2}{n_a^2+n_b^2-2 n_a n_b \cos(\vartheta)} \right|.
\label{eq:ptheta3}
\end{align}
If one of the refractive indexes is negative (Fig.~\ref{SnellTheta-crop}D),
\begin{align}
\vartheta_t=\vartheta-\vartheta_i,
\end{align}
we obtain
\begin{align}
p_{\vartheta_3}(\vartheta)=
\frac{1}{\pi} \mathbb{1}_{[0,\frac{\pi}{2} - \sign(n_a n_b) \vartheta_c]}(\vartheta)
\left|  1-\frac{n_b^2-n_a^2}{n_a^2+n_b^2-2 n_a n_b \cos(\vartheta)} \right|.
\label{eq:ptheta3n}
\end{align}
Equation~(\ref{eq:ptheta3n}) also takes into account  Eq.~(\ref{eq:ptheta3}).

By following the procedure utilized for Eq. (\ref{eq:Pt}), the probability $P_{t_{b \rightarrow a}}(\vartheta)$ to have Eq.~(\ref{eq:ptheta3}), i.e., that the photon is refracted, is
\begin{align}
P_{t_{b \rightarrow a}}(\vartheta) =
1-\frac{1}{2}\left[
\csc\left(\vartheta + 2 \arccos\left(\frac{n_b-  n_a \cos(\vartheta)}{\left(n_a^2 +n_b^2 -2 n_a n_b \cos(\vartheta)\right)^{\frac{1}{2}}}\right)\right)^2\sin(\vartheta)^2
\right.
\nonumber \\
+
\left. 
\cot\left(\vartheta + 2 \arccos\left(\frac{n_b-  n_a \cos(\vartheta)}{\left(n_a^2 +n_b^2 -2 n_a n_b \cos(\vartheta)\right)^{\frac{1}{2}}}\right)\right)^2\tan(\vartheta)^2
\right],
\label{eq:Ptba}
\end{align}
and if one of the refractive indexes is negative (Fig.~\ref{SnellTheta-crop}D),
\begin{align}
P_{t_{b \rightarrow a}}(\vartheta) 
=
1-\frac{1}{2}&\left\{
\left[\csc\left(\vartheta +2 \, \sign(n_a n_b)  \arccos\left(\frac{n_b-  n_a \cos(\vartheta)}{\left(n_a^2 +n_b^2 -2 n_a n_b \cos(\vartheta)\right)^{\frac{1}{2}}}\right)\right)\sin(\vartheta)\right]^{2 \, \sign(n_a n_b)}
\right.
\nonumber \\
&+
\left. 
\left[\cot\left(\vartheta +2 \, \sign(n_a n_b)  \arccos\left(\frac{n_b-  n_a \cos(\vartheta)}{\left(n_a^2 +n_b^2 -2 n_a n_b \cos(\vartheta)\right)^{\frac{1}{2}}}\right)\right)\tan(\vartheta)\right]^{2 \, \sign(n_a n_b)}
\right\}.
\label{eq:Ptbafinal}
\end{align}
Equation~(\ref{eq:Ptbafinal}) also includes Eq.~(\ref{eq:Ptba}) .

\subsubsection*{Probability density function   $p_{\vartheta_4}(\vartheta)$ (reflection: $b \rightarrow a$; no critical angle)}

When the photon travels from  $b \rightarrow a$ (no critical angle), the angle $\vartheta$ (as shown in Fig.~\ref{SnellTheta1-crop}) is uniformly distributed over
\begin{align}
\vartheta \in [0,\pi]
\end{align}
and thus the pdf $p_{\vartheta_4}(\vartheta )$ can immediately be written as
\begin{align}
p_{\vartheta_4}(\vartheta ) = \frac{1}{\pi } \mathbb{1}_{[0,\pi]}(\vartheta ) .
\end{align}
Once again, by inserting
\begin{align}
\vartheta_i=\vartheta_t=\frac{\pi -\vartheta}{2}
\end{align}
(Fig.~\ref{SnellTheta1-crop})  in Fresnel's formula, we obtain  
\begin{align}
P_{r_{b\rightarrow a}}(\vartheta)&=  
\frac{1}{2 n_a^2}
\left[n_a^2 + \cos(\vartheta) \left(  n_b^2-n_a^2 + n_b^2 \cos(\vartheta) \right) \right]
\nonumber \\
&\times
\cot\left( \frac{\vartheta}{2} +\arcsin\left( \left|\frac{n_b}{n_a}\right|\cos\left(\frac{\vartheta}{2}\right) \right)\right)^2
\csc\left( \frac{\vartheta}{2} +\arccos\left( \left|\frac{n_b}{n_a}\right|\cos\left(\frac{\vartheta}{2}\right)  \right)\right)^2,
\label{Eq:Prtheta1ba}
\end{align}
that holds for any positive/negative refractive index.
It is worth noting that $P_{r_{b\rightarrow a}}(\vartheta)$ [Eq.~(\ref{Eq:Prtheta1ba})] can be derived from $P_{r_{a\rightarrow b}}(\vartheta)$ 
[Eq.~(\ref{Eq:Prtheta1})] by exchanging 
$n_a$ and $n_b$.

\subsubsection*{Homogenized probability density function  $p_{\Theta}(\vartheta)$ for the isotropic medium}

Finally, the homogenized pdf $p_{\Theta}(\vartheta)$ for the  binary (isotropic-Poisson) statistical mixture
can be found by inserting Eqs. (\ref{eq:sigmasigma4}) and (\ref{eq:P0}) into Eq. (\ref{eq:pthetavattheta}) to obtain
\begin{align}
p_{\Theta}(\vartheta)&=\left(1-P \right)
\Bigl\{\frac{L\mu_{t_a}}{L\mu_{t_a}+P}p_{\rm HG}(\vartheta,g_a)
 +
\frac{P}{L\mu_{t_a}+P}
\left[P_{\rm partial}[P_{t_{a \rightarrow b}}(\vartheta)p_{\vartheta_0}(\vartheta)+ P_{r_{a \rightarrow b}}(\vartheta) p_{\vartheta_1}(\vartheta)]
+ P_{\rm total} p_{\vartheta_2}(\vartheta)\right]\Bigr\}
\nonumber \\
&+
   P
\Bigl\{\frac{L\mu_{t_b}}{L\mu_{t_b}+1-P}p_{\rm HG}(\vartheta,g_b)\Bigr.
+
\frac{1-P}{L\mu_{t_b}+1-P}
\Bigl. [P_{t_{b \rightarrow a}}(\vartheta)p_{\vartheta_3}(\vartheta)+P_{r_{b \rightarrow a}}(\vartheta)p_{\vartheta_4}(\vartheta)] \Bigr\}.
\label{eq:pdfthetafinalfinalfinal}
\end{align}
Given that $p_{\Theta}(\vartheta)$ is a pdf, we have $\int_0^{\pi}p_{\Theta}(\vartheta)d\vartheta=1$, as expected.

\subsection*{Phase function $p_{\Phi}(\varphi)$ (refraction/reflection)}

Equation (\ref{eq:pdfthetafinalfinalfinal}) represents the pdf for the polar angle $\vartheta$. 
However, in a MC simulation, we also need the azimuthal angle $\varphi$.
Due to the medium's construction, which involves using the Henyey-Greenstein pdf and the isotropic random orientation of the tessel's boundaries, $\varphi$ is uniformly distributed over
$ [0,2 \pi]$.
For this reason, the pdf for $\varphi$ can be expressed directly as
\begin{align}
	p_{\Phi}(\varphi)=\frac{1}{2\pi } \mathbb{1}_{[0,2 \pi]}(\varphi).
\end{align}

%

\subsection*{Derivation of the albedo for the random medium with mixing statistics}

To propagate a photon in a MC simulation, obeying to the pdf $p_{ s}(s;\mu_{t_i} ,\sigma_i )$ [Eq.~(\ref{eq:psFINALmixLP})], 
it is necessary to determine whether the photon is absorbed or scattered at each step. To this aim, we must calculate the homogenized albedo $\Lambda_{\rm mix}$ for  the binary (isotropic-Poisson) statistical mixture, 
as described for example in Ref. \cite{ref:Binzoni2023}, i.e.;
\begin{align}
\Lambda_{\rm mix}=
\frac{\int_0^{\Delta s} p_{s_{\rm mix}}(s;\mu_{s_a},\mu_{s_b},L,P) ds}{\int_0^{\Delta s} p_{s_{\rm mix}}(s;\mu_{t_a},\mu_{t_b},L,P) ds}
\approx
\frac{(1-P)(\mu_{s_a}+\frac{P}{L})+P(\mu_{s_b}+\frac{1-P}{L})}{(1-P)(\mu_{t_a}+\frac{P}{L})+P(\mu_{t_b}+\frac{1-P}{L})},
\label{eq:albedo}
\end{align}
where $\Delta s \ll 1$. 
Note that if $P=0$ or $P=1$ (only one type of tessel), we obtain, respectively,  the expected  classical albedos $\mu_{s_a}/\mu_{t_a}$ or $\mu_{s_b}/\mu_{t_b}$,
for homogeneous media\cite{ref:Martelli2022}.

\subsection*{MC code for the  binary (isotropic-Poisson) statistical mixture}

As explained in the Introduction section --- and as already explained in the case of a statistical mixture with matched refractive indexes\cite{ref:Binzoni2023} ---
the present approach allows us to treat the binary (isotropic-Poisson) statistical mixture
as a homogeneous medium. 
For this reason, it is sufficient to replace the look-up tables 
for
the SSF and the PF,
of any classical MC code for photon propagation, with
the look-up tables derived from the proposed Eqs.~(\ref{eq:psFINALmixLP}) and 
(\ref{eq:pdfthetafinalfinalfinal}), respectively.
The classical albedo is replaced by Eq.~(\ref{eq:albedo}).
This, with a slight difference, i.e., when the photon originates from a fixed boundary (e.g. the boundary of the homogenized medium) 
or from a fixed light source, Eq. (\ref{eq:psFINALmixLP})  must be replaced by Eq. (\ref{eq:uncorr})\cite{ref:dEon2018,ref:BinzoniMartelli2022}.
This is due to the fact that Eq. (\ref{eq:psFINALmixLP})  is not an exponential law (i.e.; it does not represent the classical Beer-Lambert-Bouguer law).
Below, we will see some explanatory example of the present homogenized MC approach. 

The look-up tables can be derived as usual, e.g., by  computing
\begin{align}
\chi_{\rm s_{mix}}  = \int_0^{s}p_{\rm s_{mix}}(s';\mu_{t_a},\mu_{t_b},L,P)ds'; 
\quad s\in [0,+\infty[,
\end{align}
\begin{align}
\chi_{\rm s_{fix}}  = \int_0^{s}p_{\rm s_{fix}}(s';\mu_{t_a},\mu_{t_b},L,P)ds'; 
\quad s\in [0,+\infty[
\end{align}
and
\begin{align}
\chi_{\Theta}  = \int_0^{\vartheta}p_{\Theta}(\vartheta')d\vartheta'; 
\quad \vartheta \in [0,\pi],
\end{align}
where $\chi_{\rm s_{mix}}$, $\chi_{\rm s_{fix}}$ and $\chi_{\Theta}$ are uniformly distributed random variables over $[0,1]$,
In practical words, the look-up tables allow  us to obtain random $s$ or $\vartheta$ as a function of  $\chi_{\rm s_{mix}}$, $\chi_{\rm s_{fix}}$ or $\chi_{\Theta}$.

\section*{Explanatory examples}
\label{sec:Explanatory examples}

\subsection*{Example 1}

\begin{figure}[htbp]
\includegraphics[width=9cm]{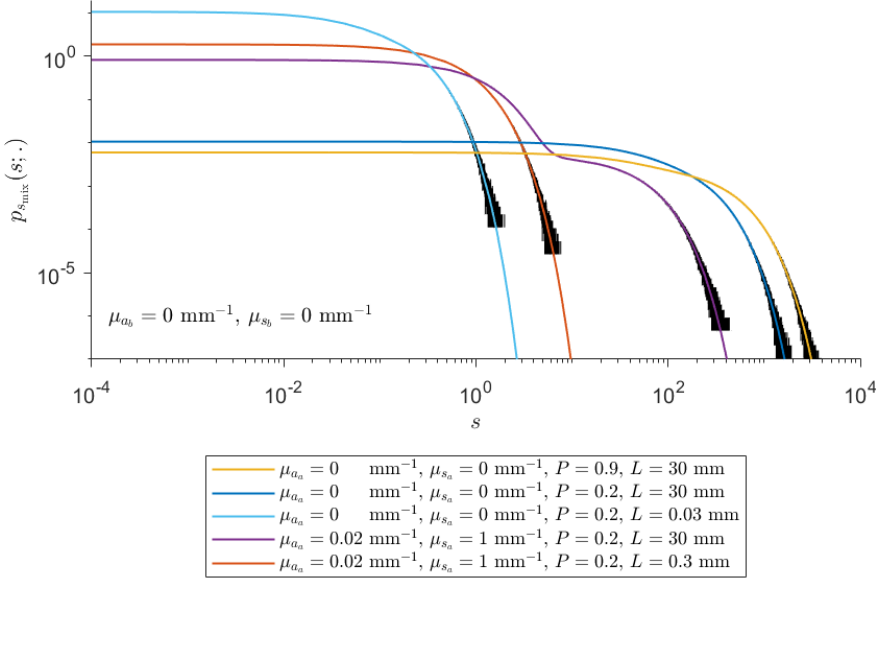}
\centering
\caption{Single step function $p_{\rm s_{mix}}(s;.)$ as a function of different optical and geometrical parameters.
The black lines represent the relative ``gold standard'' MC data.
}\label{fig_SSF-crop}
\end{figure}

\begin{figure}[htbp]
\includegraphics[width=9cm]{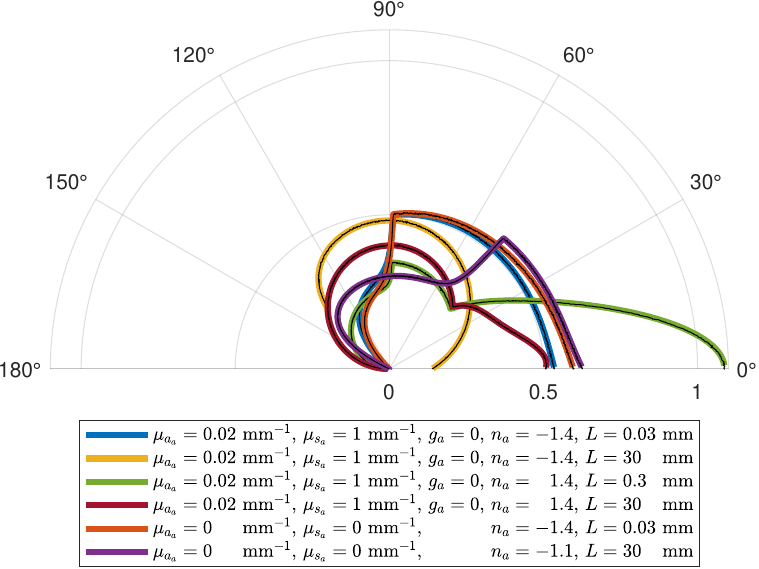}
\centering
\caption{Single step function $p_{\Theta}(\vartheta)$ as a function of different optical and geometrical parameters,
where $\mu_{a_b}=\mu_{s_b}=0$.
The black lines appearing over the colored lines represent the relative ``gold standard'' classical MC data.}\label{Fig_PF-crop}
\end{figure}

The pdfs $p_{\rm s_{mix}}(s;.)$ and $p_{\Theta}(\vartheta)$ have been tested by
developing a ``gold standard'' classical MC code, similar to that utilized 
in
 Ref\cite{ref:Binzoni2023},
 and by taking 
into account for the unmatched refractive indexes
and different tessel types.
Few examples are reported in 
Figures \ref{fig_SSF-crop} and  \ref{Fig_PF-crop},
where the black lines represent  the MC data.
A nice agreement can be observed between the ``gold standard'' classical MC data and the analytical 
(homogenized) models.

\subsection*{Example 2}

\begin{figure}[htbp]
\includegraphics[width=9cm]{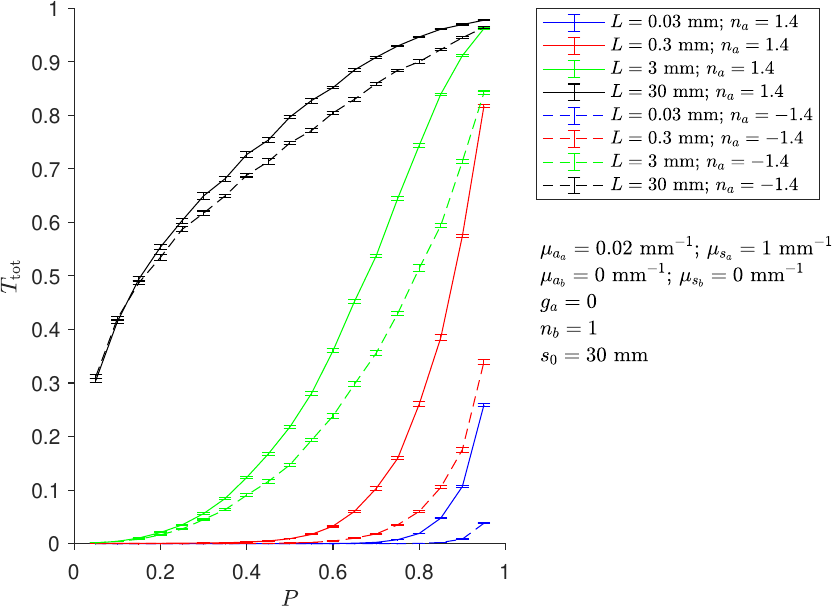}
\centering
\caption{Total transmitted light  $T_{\rm tot}$ through a semi-infinite slab of thickness $s_0$ as a function of $P$.
Vertical bars represent standard errors.}
\label{Fig_Transmission_muaa002-crop}
\end{figure}

Figure \ref{Fig_Transmission_muaa002-crop} 
represents a typical MC simulation that can be performed with
the pdfs obtained in the present contribution, i.e.; the medium 
--- binary (isotropic-Poisson) statistical mixture --- is treated as it would be ``homogeneous''.
In this example, the medium geometry is   a semi-infinite slab of thickness $s_0$. Tessels of type b are considered to be ``holes'' of  air.
The light source is a pencil beam normally impinging on one of the medium surfaces.
The total transmitted light $T_{\rm tot}$ is detected on the surface opposite to the light source.

The figure nicely reproduces the expected behavior, i.e., 
for a small $P$, the probability 1-P of having vessels of type a with non-null absorption increases.
Thus, fewer photons reach the opposite surface and  $T_{\rm tot}$ tends to 0.
As $P$ increases, $T_{\rm tot}$ also increases because the fraction of tessels of type b containing air becomes higher, making it easier for photons to pass through the slab.

The two cases for $L=30$ mm follow the same behavior, but they  go more slowly to 0 for small $P$.
It is interesting to note that, in general,  if the $n_a$ sign becomes negative while the other parameters remain the same, $T_{\rm tot}$ decreases.

\subsection*{Example 3}

\begin{figure}[htbp]
\includegraphics[width=9cm]{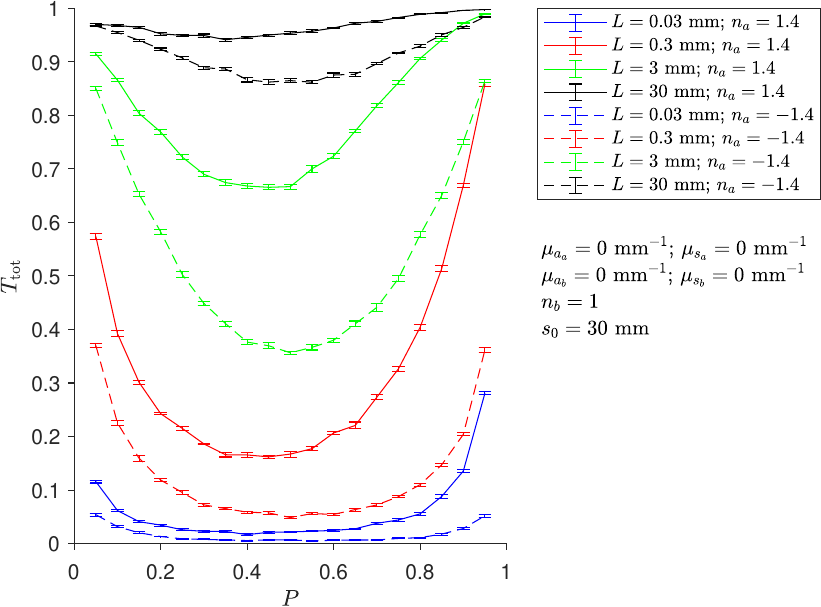}
\centering
\caption{Same MC simulation as in Example 2 section, but with $\mu_{t_a}=0$.
Vertical bars represent standard errors.}\label{Fig_Transmission_muaa0-crop}
\end{figure}

To demonstrate that the behavior of photons observed in the Example 2 section is mainly due to the absorption (and scattering) properties of type a tessels, we repeat the same MC explanatory example, but   with  $\mu_{a_a}=\mu_{s_a}=0$.
In this case, only ``scattering'' effects due to the presence of the tessels boundaries are taken into account (i.e.; no photons are absorbed; they are eliminated only by exiting the slab).
In this case, Fig. \ref{Fig_Transmission_muaa0-crop} 
nicely shows that the influence of  $\mu_{t_a}$
observed in \ref{Fig_Transmission_muaa002-crop}  disappears.
The curves tend to become symmetric around the center of the $P$ axes.

The minimum $T_{\rm tot}$ values for each curve are found for $P$ values where the probability to have tessels of type a tends to be equal to
the probability to have tessels of type b. In fact, with this condition, we have the highest density of a-to-b tessel interfaces inside the slab.
This makes it more difficult for photons to pass through the slab, i.e.; more photons are reflected on the light source side.

As in Example 2 section,  if the $n_a$ sign becomes negative, and the other parameters remain the same, $T_{\rm tot}$ decreases coherently.

\section*{Conclusions}

In the present contribution we have derived the {\it homogenized} SSF, SF and albedo for 
a binary (isotropic-Poisson) statistical mixture with unmatched refractive indexes.
The obtained mathematical tools allow us to reduce hundreds of complex MC simulations, of photon propagation in
binary (isotropic-Poisson) statistical mixture, to only one simpler MC simulation of an equivalent homogeneous medium.
This homogenized MC simulation is not an approximation, but is a mathematically equivalent approach, 
built analytically, giving the same results
as the classical MC.

The tutorial examples presented in the last section already show the potential interest that can be generated by studying the influence of negative refractive indexes on photon propagation, or the influence of tessel geometry on photon scattering behavior.
To the best of our knowledge, these findings have never been reported elsewhere.
We hope that these  preliminary results will stimulate future investigations 
in  new domains related to binary (isotropic-Poisson) statistical mixtures.


\section*{Author contributions statement}

TB and AM contributed equally to this work. All authors reviewed the manuscript.

\section*{Data availability}

The datasets used and/or analysed during the current study available from the corresponding author on reasonable request.

\section*{Additional information}

\textbf{Competing financial interests:} The authors declare no competing financial interests.

\end{document}